\documentclass[manuscript]{acmart}
\usepackage{caption}
\usepackage{subcaption}
\usepackage{amsmath}
\usepackage{mathtools}
\usepackage{graphics}
\AtBeginDocument{%
  \providecommand\BibTeX{{%
    \normalfont B\kern-0.5em{\scshape i\kern-0.25em b}\kern-0.8em\TeX}}}

\copyrightyear{2022}
\acmYear{2022}
\setcopyright{rightsretained}
\acmConference[UbiComp/ISWC '22 Adjunct]{Proceedings of the 2022 ACM International Joint Conference on Pervasive and Ubiquitous Computing}{September 11--15, 2022}{Cambridge, United Kingdom}
\acmBooktitle{Proceedings of the 2022 ACM International Joint Conference on Pervasive and Ubiquitous Computing (UbiComp/ISWC '22 Adjunct), September 11--15, 2022, Cambridge, United Kingdom}\acmDOI{10.1145/3544793.3560360}
\acmISBN{978-1-4503-9423-9/22/09}
\begin{document}

%%
%% The "title" command has an optional parameter,
%% allowing the author to define a "short title" to be used in page headers.
\title{You Are What You Use: Usage-based Profiling in IoT Environments}

%%
%% The "author" command and its associated commands are used to define
%% the authors and their affiliations.
%% Of note is the shared affiliation of the first two authors, and the
%% "authornote" and "authornotemark" commands
%% used to denote shared contribution to the research.

\author{Manan Choksi}
\email{mcho6881@uni.sydney.edu.au}
\orcid{0000-0002-5990-3867}
\affiliation{%
  \institution{The University of Sydney}
  \city{Sydney}
  \state{NSW}
  \country{Australia}
  \postcode{2000}
}

\author{Dipankar Chaki}
\email{dipankar.chaki@sydney.edu.au}
\orcid{0000-0002-4048-8798}
\affiliation{%
  \institution{The University of Sydney}
  \city{Sydney}
  \state{NSW}
  \country{Australia}
  \postcode{2000}
}

\author{Abdallah Lakhdari}
\email{abdallah.lakhdari@sydney.edu.au}
\orcid{0000-0001-8005-1534}
\affiliation{%
  \institution{The University of Sydney}
  \city{Sydney}
  \state{NSW}
  \country{Australia}
  \postcode{2000}
}

\author{Athman Bouguettaya}
\email{athman.bouguettaya@sydney.edu.au}
\orcid{0000-0003-1254-8092}
\affiliation{%
  \institution{The University of Sydney}
  \city{Sydney}
  \state{NSW}
  \country{Australia}
  \postcode{2000}
}

%%
%% By default, the full list of authors will be used in the page
%% headers. Often, this list is too long, and will overlap
%% other information printed in the page headers. This command allows
%% the author to define a more concise list
%% of authors' names for this purpose.
\renewcommand{\shortauthors}{Manan and Dipankar, et al.}

%%
%% The abstract is a short summary of the work to be presented in the
%% article.
\begin{abstract}
Habit extraction is essential to automate services and provide appliance usage insights in the smart home environment. However, habit extraction comes with plenty of challenges in viewing typical start and end times for particular activities. This paper introduces a novel way of identifying habits using an ensemble of unsupervised clustering techniques. We use different clustering algorithms to extract habits based on how static or dynamic they are. Silhouette coefficients and a novel noise metric are utilized to extract habits appropriately. Furthermore, we associate the extracted habits with time intervals and a confidence score to denote how confident we are that a habit is likely to occur at that time.
\end{abstract}

\vspace{-5mm}

%%
%% The code below is generated by the tool at http://dl.acm.org/ccs.cfm.
%% Please copy and paste the code instead of the example below.
%%


\vspace{-5mm}

%%
%% Keywords. The author(s) should pick words that accurately describe
%% the work being presented. Separate the keywords with commas.

%% A "teaser" image appears between the author and affiliation
%% information and the body of the document, and typically spans the
%% page.
% \begin{teaserfigure}
%   \includegraphics[width=\textwidth]{sampleteaser}
%   \caption{Seattle Mariners at Spring Training, 2010.}
%   \Description{Enjoying the baseball game from the third-base
%   seats. Ichiro Suzuki preparing to bat.}
%   \label{fig:teaser}
% \end{teaserfigure}

%%
%% This command processes the author and affiliation and title
%% information and builds the first part of the formatted document.
\maketitle
%\vspace{-5mm}
\section{Introduction}
Today IoT is one of the most ubiquitous terms used in tech as its value and popularity have risen over the past decade. IoT refers to the web of interconnected devices that can provide insights and automate many services. Given its popularity, concepts like smart homes and smart campuses are becoming pioneering applications in this domain. Smart homes help provide users with detailed information about their energy consumption while providing avenues to automate services like turning devices on in response to known user habits. To provide this level of convenience, we need to develop a sophisticated technique to detect user habits. A habit refers to a set routine or task a user performs at a particular time. Examples of habits include eating breakfast, watching television, showering, or exercising. We need to discern certain activities' start and end times to detect these patterns and analyze habits. Despite various techniques to extract habits, methods related to identifying definite habit start and end times are mainly unaddressed. Many papers rely primarily on pre-setting arbitrary time bounds for certain activities or using time bounds that are not precise \cite{banaee2020explaining, wang2016new, chaki2020fine}. We believe that identifying precise time intervals with some confidence level can be much more helpful if we want to use habit extraction methods for developing personalized services in the future. As a result, this paper aims to introduce a novel way of identifying habits using an ensemble of unsupervised clustering techniques to cluster activities' start and end times. Since some user behaviors may be more static than others, we use different clustering algorithms like k-means, hierarchical clustering, and density-based clustering to profile different usage behaviors of people. For example, one may predict that habits like eating breakfast are more likely to occur during the morning. However, habits like watching television are more prone to be random (i.e., people may watch television in the morning, afternoon, or evening). This makes it essential to employ different clustering techniques to capture different usage behavior patterns.

\begin{figure}
%\vspace{-5mm}
\centering
    \setlength{\abovecaptionskip}{1pt}
    \setlength{\belowcaptionskip}{-5pt}
\includegraphics[width=.8\linewidth]{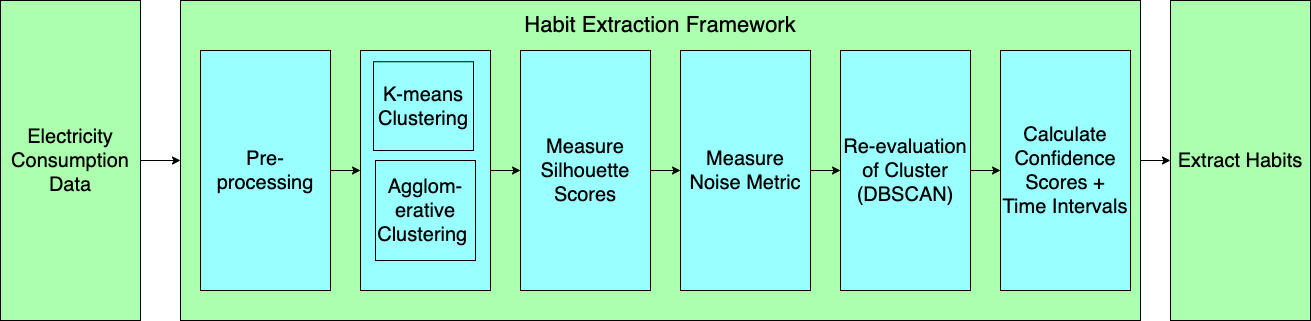}
\caption{Proposed Habit Extraction Framework.}
\vspace{-5mm}
\end{figure}

\vspace{-3mm}

\section{Proposed Methodology}

%\vspace{-5mm}

\subsection{K-means and Agglomerative Clustering}
\vspace{-2mm}
We aim to build a model to profile typical start and end times for particular user habits. To do this, we systematically implement different clustering algorithms and settle on a clustering pattern in accordance with two metrics: silhouette coefficient and a novel noise metric. To begin, we pass our tuples denoted by ($t_{start}$, $t_{end}$) for one activity to be clustered by the k-means and agglomerative clustering algorithms. Then, we denote the number of clusters as $k$ and run both algorithms from $k$ equals 2 to $k$ equals any maximum value that can be specified arbitrarily by the user, increasing the value of $k$ by one and calculating the silhouette score each time. Finally, we pick the clustering arrangement with the highest silhouette score (closest to 1) as it represents the most compact and well-separated clusters.

\vspace{-4mm}

\subsection{Novel Noise Metric}
\vspace{-2mm}
Once we find the optimal number of clusters and the favored clustering technique using this method, we propose a noise metric to measure the sparsity of each cluster. First, we find the euclidean distance between every combination of two tuples in every cluster generated. We can denote the distance between any two tuples as $d(tup_{1}, tup_{2})$ where $tup_{n}$ represents a single tuple for a start and end time of activity in a day. Then, for each cluster, we find the mean of all $\frac{n(n - 1)}{n}$ distances generated where $n$ represents the number of tuples in a cluster. We can denote this mean value as $P_{r}$ where $r$ uniquely identifies each generated cluster.

\vspace{-7mm}
\begin{equation}
%\vspace{-2mm}
    P_r = \frac{\sum\limits_{0\leq i<j\leq n} ^ {n}d(tup_i,tup_j)}{n}
%\vspace{-2mm}
\end{equation}
% \[
%     P_r = \frac{\sum\limits_{0\leq i<j\leq n} ^ {n}d(tup_i,tup_j)}{n}
% \]

%At this point, somebody who wishes to use our model must define a threshold that $P_{r}$ for any cluster must not exceed for it to be a valid clustering that is compact and free from excessive amounts of 'noise'. Their threshold represents how granular and precise they want specific habit start and end times to be.

\vspace{-6mm}

\subsection{DBSCAN}
\vspace{-2mm}
Given that the value $P_{r}$ generated for all clusters falls below a certain threshold, we can finish our process and calculate our results. On the contrary, if any $P_{r}$ value for a cluster exceeds a set threshold, the clusters display too much variability. In this case, we turn to a density-based clustering method called DBSCAN. Before we perform DBSCAN, we must decide on an $\epsilon$ value that specifies the maximum distance between two points in one cluster. On top of this, we must specify any number $v$, which denotes the minimum number of points to form a cluster. We can set the value for $\epsilon$ using the elbow method in DBSCAN. After performing DBSCAN, we can test the validity of the clusters by calculating their $P_{r}$ values. If they still exceed our thresholds, we can reduce our $\epsilon$ value, but if not, we can proceed to extract habits.

\vspace{-5mm}

\subsection{Habit Extraction}
\vspace{-2mm}
In the habit extraction process, we calculate the average of every single start and end time in a singular cluster. After this, we calculate the standard deviation of the start and end times in each cluster to associate possible times an activity could begin. Furthermore, we will associate a confidence score as a probability to represent how often we believe the user will carry out the activity at the clustered time interval. We can represent this as a probability using $\frac{q}{n}$, where $q$ denotes the total number of tuples in a cluster, and $n$ represents the total number of tuples for the entire activity across all clusters.\looseness=-1

\vspace{-3mm}

\section{Experiments}
\vspace{-2mm}
%\subsection{Data Preparation}
The majority of the data used to test the methods in this paper come from the University of Washington's CASAS household data set and REFIT electrical loads management data set \cite{cook2012casas,murray2017electrical}. A significant difference between the data sets was their presentation of ON and OFF states. While CASAS pre-processed the voltage data to signify ON and OFF states, REFIT required analyzing the voltage readings and determining a threshold to represent an ON or OFF state for the device. Based on the ON times for devices presented in the data, we represented how long a device/activity was ON for as a two-dimensional tuple ($t_{start}$, $t_{end}$) where $t_{start}$ is the start time for a device and $t_{end}$ is the end time.

\begin{figure}%
%\vspace{-5mm}
    \centering
    \setlength{\abovecaptionskip}{7pt}
    \setlength{\belowcaptionskip}{-10pt}
    \subfloat[\centering]{{\includegraphics[width= 4cm]{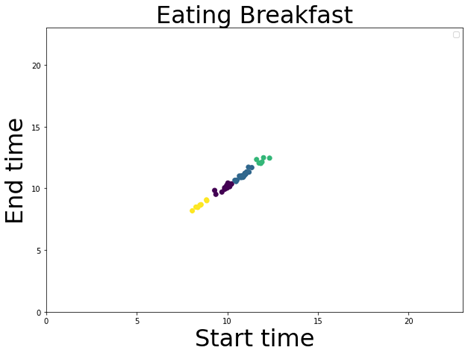} }}%
    \qquad
    \subfloat[\centering]{{\includegraphics[width=4cm]{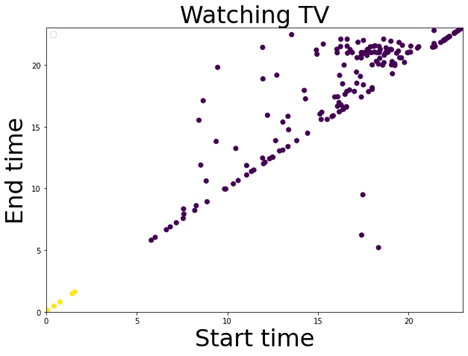} }}%
    \qquad
    \subfloat[\centering]{{\includegraphics[width=4cm]{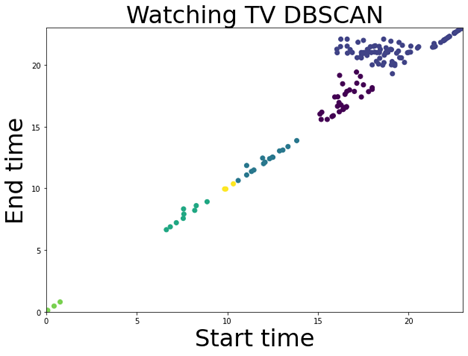} }}%
    %\vspace{-3mm}
    \caption{Cluster representation of different habits: (a) hierarchical clustering of breakfast eating habits, (b) k-means clustering of television watching habits, and (c) DBSCAN clustering of television watching habits.}%
    \label{fig:example}%
    %\vspace{-3mm}
\end{figure}

In the results of our experiments, we found that k-means and hierarchical clustering were optimal in profiling usage behaviors for more regular activities. For example, figure 2a represents the data clustering to do with eating breakfast. We obtained four common times for this user to eat breakfast: 8:30am $\pm$ 18 minutes - 8:38am (18\% confidence), 9:52am $\pm$ 18 minutes - 10:03 am (24\% confidence), 10:53am $\pm$ 14 minutes - 11:04am (44\% confidence) and 11:54 am $\pm$ 14 minutes - 12:14am (13\% confidence) . On the other hand, hierarchical clustering on the television watching habits of a user suggested that common television watching times were 5:51 pm $\pm$ 276 minutes - 7:02 pm $\pm$ 285 minutes (97\% confidence) and 12:32 am $\pm$ 37 minutes - 12:37 am $\pm$ 37 minutes (3\% confidence). The results for the clustering of this activity can be viewed in figure 2b. The clusters in figure 2b cannot be viable because our noise metric score of 7.46 far exceeded our threshold for $P_{r}$, which was 4. Figure 2b represents a prime example of the necessity of introducing our noise metric. While we believe the silhouette score works well most of the time, a major component of the score lies in how well separated the data is. While this is important in measuring the feasibility of clusters, since our data do not possess high dimensionality and work on a relatively small axes size (24-hour time), we need a measure to quantify the compactness of clusters to support the silhouette score. Thus, our noise metric can measure how distinct clusters are and help detect when either hierarchical or k-means clustering struggles to output viable clusters due to the sparsity of data. Figure 2c shows the results of DBSCAN on the usage data from figure 2b. As we can see, many of the noisy points have been pruned, and the clustering generated is far more compact and precise, which is supported by the fact that all 5 clusters for figure 2c possessed a noise metric that fell below our threshold of 4.

\vspace{-3mm}

\section{Conclusion and Future Work}

\vspace{-2mm}

This paper proposes a method to extract habits' start and end times using a group of unsupervised clustering algorithms. Furthermore, we introduce a novel noise metric to address the deficiencies related to measuring the fitness of clusters. In addition, we associate extracted habits with some confidence scores and upper and lower time limits. In the future, we plan to use our methods to develop a more extensive framework for detecting changes in habits.

\vspace{-3mm}

\section*{Acknowledgment}
\vspace{-2mm}
This research was partly made possible by  LE220100078 and LE180100158 grants from the Australian Research Council. The statements made herein are solely the responsibility of the authors.

\bibliographystyle{ACM-Reference-Format}
\bibliography{sample-base}

\end{document}